\documentclass[epj]{svjour}
\usepackage{graphics}
\usepackage{epstopdf}

\begin{document}
\title{Dynamical prediction of flu seasonality driven by ambient temperature: influenza {\it vs.} common cold}
\author{E.B. Postnikov}                     
\institute{Department of Theoretical Physics, Kursk State University,
Radishcheva st., 33, 305000, Kursk, Russia \email{postnicov@gmail.com}}
\date{Received: date / Revised version: date}
%
\abstract{
This work presents a comparative analysis of {\tt Influenzanet} data for influenza itself and common cold in the Netherlands during the last 5 years, from the point of view of modelling  by linearised SIRS equations parametrically driven by the ambient temperature. It is argued that this approach allows for the forecast of common cold, but not of influenza in a strict sense. The difference in their kinetic models is discussed with reference to the clinical background. 
} 
\maketitle
\section{Introduction}

Influenza-like diseases are most common kind of seasonal illness in regions of moderate climate, the typical example of which is Europe. At the same time, even the common cold  may result in more serious complications, such as pneumonia, pleurisy, sinusitis, ear infections, etc. \cite{Heikkinen2003}. Another example of such diseases is influenza itself, the occurrence of which is associated with both seasonality and abrupt outbreaks \cite{Cannell2008}. The considerable impact of influenza-like diseases on public health systems has  resulted in the active development of flu monitoring systems based on various approaches. Amongst the basic sources of data for Europe, one can list the {\tt European Influenza Surveillance Network} (EISN) \cite{EISN} based on the reports of sentinel general practitioners, {\tt Google Flu Trends} \cite{Gflu} based on the analysis of certain search terms in requests to Google, and {\tt Influenzanet}, a system, which monitors the activity of influenza-like-illness (ILI) and common cold using volunteers reports via the Internet \cite{InflNet}.

The availability of such data allows researchers to seek a solution to the more important problem: how to  predict a possible epidemics in advance. Recently, there have been various approaches to this problem and a comprehensive review of the state of the art can be found in \cite{Chretien2014}.  The correlated synchronisation between variations of the ambient temperature and activity of the common cold was revealed in the pioneering work by van~Loghem \cite{vanLoghem1928}, and confirmed by further extensive studies \cite{Fellowes1973,Mourtzoukou2007,Meerhoff2009}.  Moreover, the interplay between weather conditions and the epidemic level of respiratory diseases has the definite medical basis  \cite{Eccles2002,Pica2012,Lowen2014}

As recent successful studies related to influenza-like diseases, the works \cite{Shaman2012,vanNoort2012,Postnikov2013} may be referenced. The authors of the first of these works consider the humidity-forced susceptible-–infectious–-recovered–-susceptible (SIRS) mathematical model combining both dynamical and stochastic calculations, which have been practically realised for the influenza forecasts in the USA during the 2012-2013 season \cite{Shaman2013}. 

The study described in  \cite{vanNoort2012} considers a distinction between influenza itself and influenza-like illnesses in Europe, based on the SIR model driven by both the ambient temperature and the humidity using {\tt Influenzanet} data. However, the SIR model does not include recovery into the susceptible state again, therefore, a manual reset of the initial conditions was introduced for each season separately. On the other hand, the results presented in article \cite{Postnikov2013} show that the temperature-driven SIRS model allows for generation of long-term prognoses using the set of constant parameters defined for a given geographical location. This model of non-linear differential equations has been tested via the examples of two European cities with the help of {\tt Google Flu Trends} data, which does not distinguish influenza-like diseases, i.e. influenza and common cold. 

Thus, the main goals of the present work are as follows: to analyse an application of the last approach using the more representative data of {\tt Influenzanet}, to reveal which kinds of seasonal diseases could be modelled, and to present the details of the dynamical system responsible for this description.

\section{Data and models}

The modelled data is taken from the {\tt Influenzanet} system \cite{InflNet}, which monitors the epidemiological situation in several European countries. The dataset from the Netherlands is one of the most representative; thus, this work deals exclusively with  this dataset.
In addition, it is worth noting the subdivision of the flu outbreak data with respect to the kind of influenza-like illness: influenza or common cold. Accordingly to \cite{InflNet}, influenza (or influenza-like disease) is detected for the set symptoms: fever or feverishness (chills), malaise, headache (at least one of them), muscle pain and simultaneously cough, sore throat, shortness of breath (at least one of them). On the other hand, the common cold is characterised by at least two symptoms: runny or blocked nose, sneezing, cough, or sore throat but without the more complicated symptoms. From the microbiological point of view, influenza is caused by influenza type A and
B viruses, which changes slowly with respect to their
surface antigens, haemagglutinin (H) and neuraminidase (N) \cite{Hampson2006}. The common cold is caused by a large variety of  rhinoviruses, respiratory
syncytial viruses (RSV), etc.  It is worth pointing pointed out the absence of a suitable common antigen across them and the significant dependence on individual resistance and susceptibility based on a person's physical conditions, including physiological and even psychological stresses  \cite{Heikkinen2003}. 

\begin{figure}
\resizebox{0.48\textwidth}{!}{\includegraphics{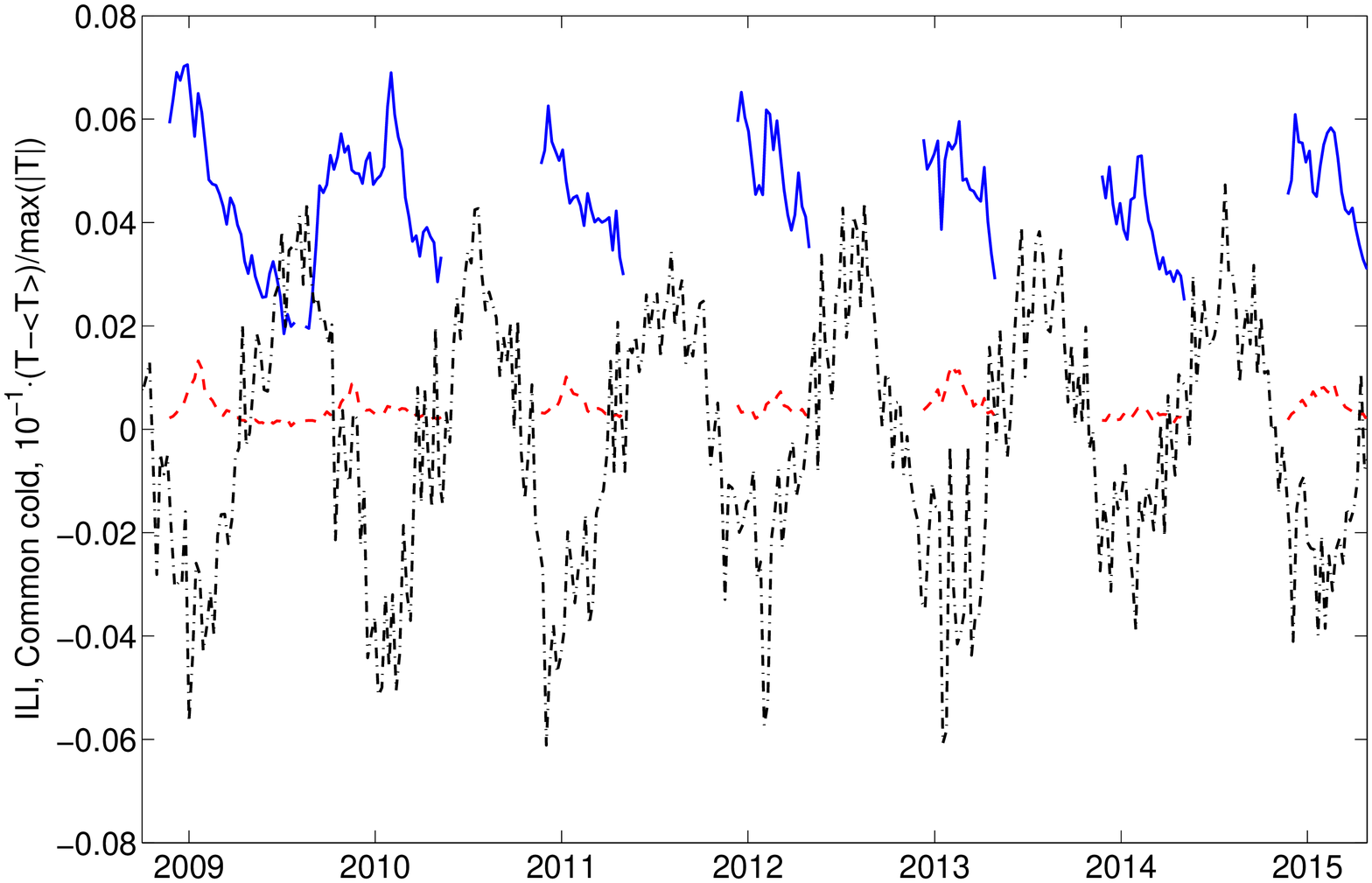}}
\caption{(Colour online) The dynamics of influenza (dashed line, red online), common cold (solid line, blue online) and the averaged ambient temperature (dash-dotted line, black online). The disease data represent the percentage of ill respondents in the survey. The dimensionless temperature curve shows the scaled relative deviation from the mean value over the considered time range.}
\label{datacurves}
\end{figure}

The daily mean temperature data series are taken from {\tt European Climate Assessment \& Dataset} \cite{eca,KleinTank2002}. Since the epidemic data are presented for the Netherlands as the whole, the mean daily temperature curves are averaged over all data from the meteorological stations located between $3.5^o$~E--$7^o$~E and $51^o$~N--$54^o$~N. These three curves are represented in Fig.~\ref{datacurves}. One can see a definite anticorrelation between maxima and minima of the temperature and the diseases data. 

Therefore, our interpretation of the data is based on the approach proposed in the work described in \cite{Postnikov2013}, using the SIRS \\ (Susceptible--Infected--Recovered--Susceptible) model: 
\begin{eqnarray}
\frac{dS}{dt}&=&-k(T(t))IS+\theta^{-1}R,\label{eqS} \\
\frac{dI}{dt}&=&k(T(t))IS-\tau^{-1}I,\label{eqI}\\
\frac{dR}{dt}&=&\tau^{-1}I-\theta^{-1}R \label{eqR}
\end{eqnarray}
with constant characteristic durations of the illness $\tau$ and  loss of immunity $\theta$. The parameter $k(T(t))$ is a function of the time-varying ambient temperature $T(t)$.

Note that it is reasonable to refer to $k(T(t))$ as ``the probability of catching flu'' in a general sense instead of the pure ``contact rate'' for the reasons mentioned above for the common cold. It will be discussed below in details that the last term, which is commonly used and completely valid for diseases such as flu (influenza in a  strict sense), measles, plague, etc., does not precisely reflect the situation for the common cold.

The system (\ref{eqS})--(\ref{eqR}) has two definite stationary states for $k=k_0=\mathrm{const}$: the unstable state $I_{s0}=0$, $R_{s0}=0$, $S_{s0}=1$ and the stable state:
\begin{equation}
S_s=\frac{1}{k\tau},\,R_s=\frac{1-\tau^{-1}k^{-1}_0}{1+\tau\theta^{-1}},\, I_s=\frac{\tau}{\theta}R_s.
\label{sstates}
\end{equation}

As for a time-dependent $k(t)$, the more general expression $k=k_0\left[1+\kappa\left(T(t-\Delta)\right)\right]$, which includes the time delay $\Delta$, will be considered in contrast to the work described in \cite{Postnikov2013}. 

We emphasise, however that the presence of $\Delta$ does not affect the reduction of the system (\ref{eqS})--(\ref{eqR}) to the form of ordinary differential equations with an explicit time-dependent outer excitation \cite{Postnikov2013}:
\begin{eqnarray}
\frac{dr}{dt}&=&N, \label{eqr}\\
\frac{dN}{dt}&=&
R_s\theta^{-1}\left(k-\tau^{-1}-k\left[1+\tau\theta^{-1}\right]R_s\right)-\nonumber\\
&&-\left(\tau^{-1}+\theta^{-1}+R_sk\left[1+2\tau\theta^{-1}\right]-k\right)N-\nonumber\\
&&-\theta^{-1}\left(\tau^{-1}+2R_sk\left[1+\tau\theta^{-1}\right]-k\right)r- \label{eqn}\\
&&-k\tau N^2-k(1+\tau\theta^{-1})Nr-k\theta^{-1}\left[1+\tau\theta^{-1}\right]r^2, \nonumber
\end{eqnarray}
obtained via the variable alteration $N=\tau^{-1}I-\theta^{-1}R$, $r=R-R_s$ (the conservation law $S+I+R=1$ is taken into account).  

This $(r,N)$ representation of the SIRS equations in the explicit form of an inhomogeneous ODE system shows that the variable $k(T(t))$ provides leading excitation as a direct outer forcing.

Fig.~\ref{datacurves} shows that $\mathrm{max}(I(t))<<1$ for both kinds of diseases. Thus, the oscillations have a small magnitude and we can linearise the system (\ref{eqr})--(\ref{eqn}) around its steady state. 

The result is represented as a simple second-order ODE:
\begin{equation}
\frac{d^2r}{dt^2}+\lambda\frac{dr}{dt}+\omega_0^2r=R_s\theta^{-1}\tau^{-1}\kappa\left(T(t-\Delta)\right)
\label{eqrlin}
\end{equation}
with positive constants $\omega_0^2=\theta^{-1}\left(k_0-\tau^{-1}\right)$, $\lambda=\tau^{-1}+\theta^{-1}+R_sk_0\left[1+2\tau\theta^{-1}\right]-k_0$. 

Due to the linearity of this equation, its solution could be considered simply as a mapping of $\kappa\left(T(t)\right)$ into $I(t)$ using the Green function method. 

The inverse transformation $I=\tau^{-1}\theta(R_s+r)+\tau\dot{r}$ results in the direct mapping given by the convolution: 
\begin{equation}
I(t)=I_s+\frac{R_S}{\theta}\int_{\Delta}^t
\kappa(t'-\Delta)G(t-t')dt',
\label{convI}
\end{equation}
where:
$$
G(\xi)=\frac{1}{\omega}e^{-\frac{\lambda}{2}\xi}\left[\textstyle{\left(\theta^{-1}-\frac{\lambda}{2}\right)}\sin(\omega\xi)+\omega\cos(\omega\xi)\right]
$$
is the Green function. 

At the same time, the principal question is the applicability of this dynamical approach to the particular kinds of respiratory diseases  under consideration. This question can only be answered by comparison with the experimental data.

\section{Results}
The constants used for the solution of Eq.~(\ref{convI}) are $k_0=0.24$, $\kappa=-0.08$, $\tau=7$~(days), $\theta=28$~(days), $\Delta=3$~(days). These values of the parameters provide the  value of the basic reproduction number $\mathbf{R_0}=(k_0\pm\kappa)\tau=1.68\pm0.56$. It overlaps with the range typical for known studies of the common cold induced by RSV ($1.2-2.1$) \cite{Weber2001} as for seasonal influenza ($0.9-2.1$) \cite{Chowell2008}, ($1.6-3$) \cite{Truscott2011}. Note also that the values of this characteristic parameter are close for both kinds of seasonal diseases.

\subsection{Influenza}

The analysis starts with the approximation of influenza-like diseases (ILI) available from {\tt Influenzanet} for the period of observations 2009-2015. 

Since the normalisations for the model and for the observations are not  directly balanced initially, the scaling and shift procedure $\alpha I+\beta\to I$ should be evaluated using some reference time range. The full season of 2009  is most suitable for this aim, because of the fullest representation of the observational data including minima of activity. 

Note that such a procedure does not violate the assumed conservation law in principle. The form $S+I+R=1$ itself is obtained by the normalisation $S/P\to S$, $I/P\to I$, $R/P\to P$ from $S+I+R=P$, where $P$ is the total considered population. However, the value of $P$ cannot be defined as an exact number. In particular, one can not simply take the country's population, since some people may not be included in surveys, may have good health conditions excluding them from the seasonal respiratory epidemic processes, etc. In fact, it is a known problem of uncertainty quantification in epidemiological modelling, and the adjustment to observed data for some referent time interval is one of the most widely accepted approaches \cite{oneill2010}. At the same time, the scaling and shift procedure $\alpha I+\beta\to I$ is linear as well as the the conservation law, which contains the linear combination of the variables. As a results, the linearity of the conservation law is not violated after this transformation.

\begin{figure}
\resizebox{0.49\textwidth}{!}{\includegraphics{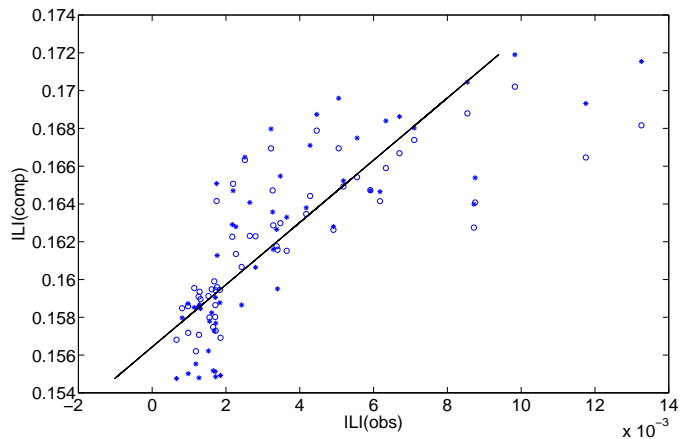}}
\caption{The plot of the calculated influenza activity (non-scaled) {\it vs.} the observational activity for the season of 2009. Circles and asterisks mark the usage of the linearised (\ref{convI}) and the full non-linear (\ref{eqS})--(\ref{eqR}) models, respectively. The solid line is the linear fit.}
\label{iliscale}
\end{figure}

Fig.~\ref{iliscale} demonstrates the aforementioned correspondence between the computed and the observed values. To control the availability of the linear equation, the results of additional calculations, which  utilise the full non-linear ODE system, are added to Fig.~\ref{iliscale}. One can see that both data clouds overlap each other in general, i.e. the simplified time series may be discussed. 

\begin{figure}
\resizebox{0.49\textwidth}{!}{\includegraphics{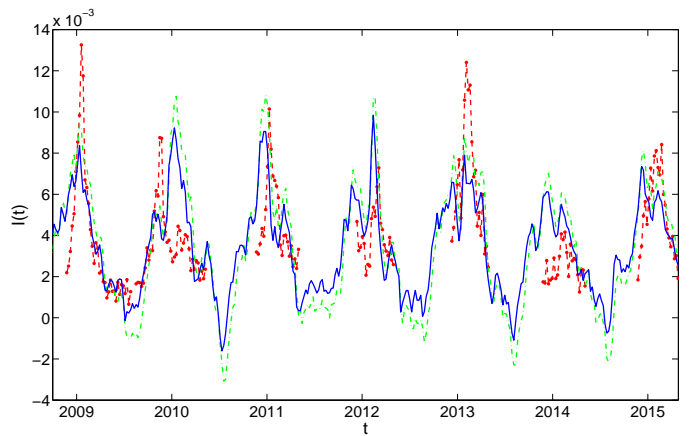}}
\caption{(Colour online) Comparison of {\tt Influenzanet}'s survey data on influenza for the Netherlands (red curves) with the prediction obtained via solution of the linearised SIRS equations, which map the ambient air temperature over the same territory into the number of infected persons (blue curve, scaled to fit observations).}
\label{ilifig}
\end{figure}

One can see that the points are highly scatted, although a definite trend presents visibly. 
Due to the large scattering of the data, the robust non-parametric method of the scale parameter regression \cite{Postnikov2015} is applied. The  parameters  obtained are: $\alpha=0.607$, $\beta=-0.095$, and the linear trend line is drawn in Fig.~\ref{iliscale} as well. These parameters are used to plot the model curve in Fig.~{\ref{ilifig}} (solid line).

One can see that the yearly oscillatory behaviour is qualitatively reproduced, but that the short time scale fluctuations visibly deviate from the observational data. Although there is some qualitative correspondence between the locations of spikes, their magnitudes differ. Moreover, the most significant failing is the negative values of the scaled simulated curve which are is completely unrealistic. Consideration of the full ODE system (\ref{eqS})--(\ref{eqR}) does not save the situation (the dash-dotted curve in Fig.~{\ref{ilifig}). The negative values could be avoided with another scaling shift, but this approach results a significantly larger difference between the details of the simulated and observed time series. 

Thus, we conclude that ambient temperature driving can not be considered as the primary factor underlying ILI epidemic dynamics. It is one of the governing factors, but one needs to consider other factors (variation of the contact rate, social dynamics in general, etc.) for an accurate forecast. 

\subsection{Common cold}

Consider now the comparison of simulations with the dynamics of common cold presented by {\tt Influenzanet}. As before, the season 2009 is used as a reference for the scaling, see Fig.~\ref{coldscale}. One can see that the data cloud is practically symmetric and squeezed around the line of linear fit with parameters $\alpha=3.881$, $\beta=-0.586$. 

\begin{figure}
\resizebox{0.49\textwidth}{!}{\includegraphics{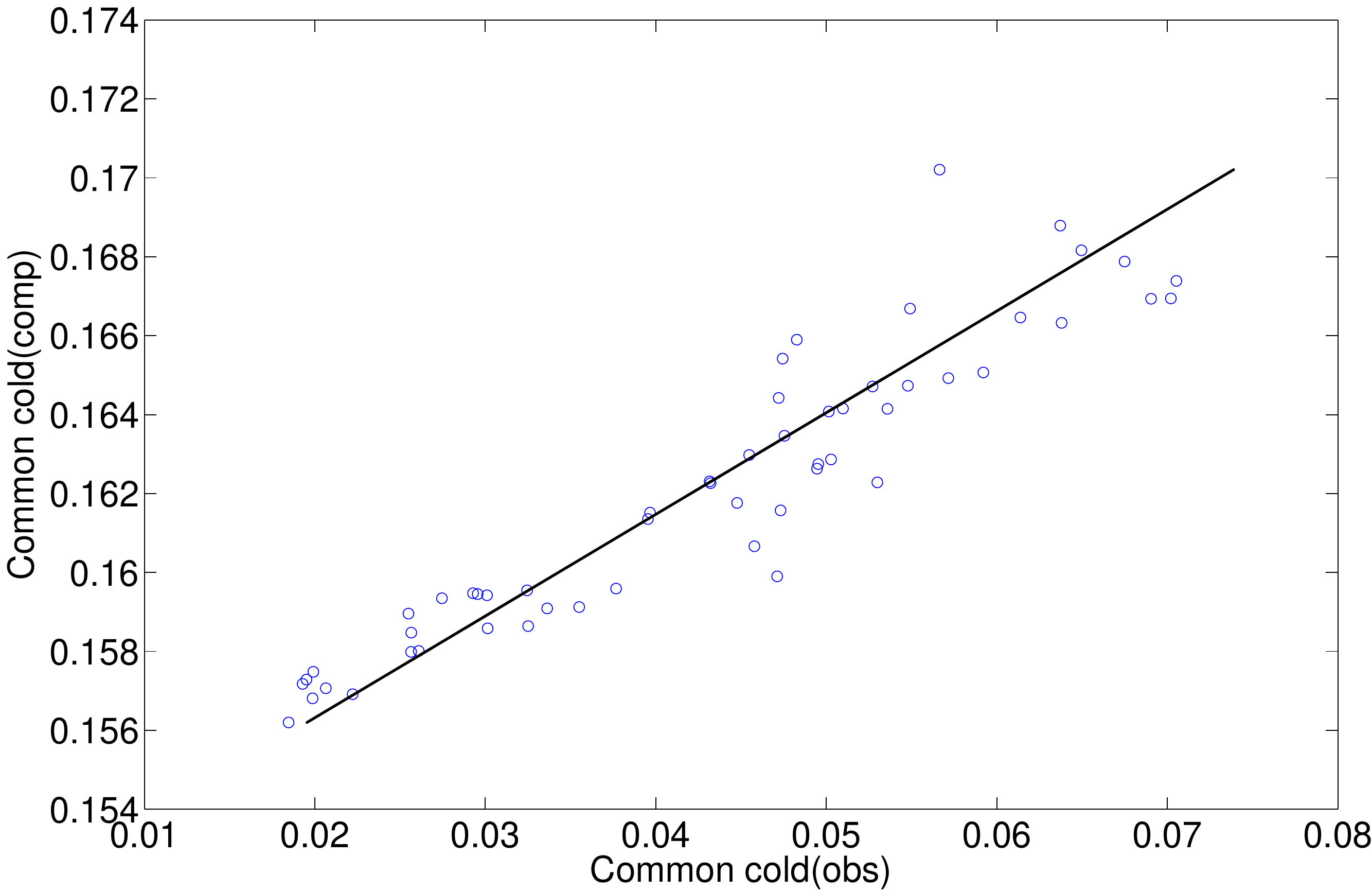}}
\caption{The plot of the calculated common cold activity (non-scaled) {\it vs.} observational activity for the season of 2009. The solid line is the linear fit.}
\label{coldscale}
\end{figure}

Correspondingly, the rescaled simulated curve is shown in Fig.~\ref{ccoldfig} as the solid line. One can see quite good correspondence with the dashed line, which presents observation data for the large and small scale time dynamics as well as for the magnitudes of a majority of outbreaks. Naturally, they coincide especially well for the interval used for the scaling, and the calculated correlation coefficient is equal to $0.94$ (without delay it is equal to $0.92$ and the shape correspondence is worse; larger delays affect the shape coincidence initially and diminish the correlation further). A special feature is the bimodal or trimodal character of some maxima of the common cold outbreaks. The calculated model based on the ambient temperature variations captures these features; in particular, see maxima for the winter seasons 2010-2015. Consideration of the ambient temperature curve, the dash-dotted line in Fig.~\ref{datacurves}, indicates that there are several short thaws for these time intervals. Thus, the harmonic filter in (\ref{convI}) catches and smooths This suggests that, in contrast to ILI, the epidemiological dynamics of the common cold are primarily based on the ambient temperature variations and can be described via the corresponding mathematical model, which takes into account their instant dynamics.

\begin{figure}
\resizebox{0.49\textwidth}{!}{\includegraphics{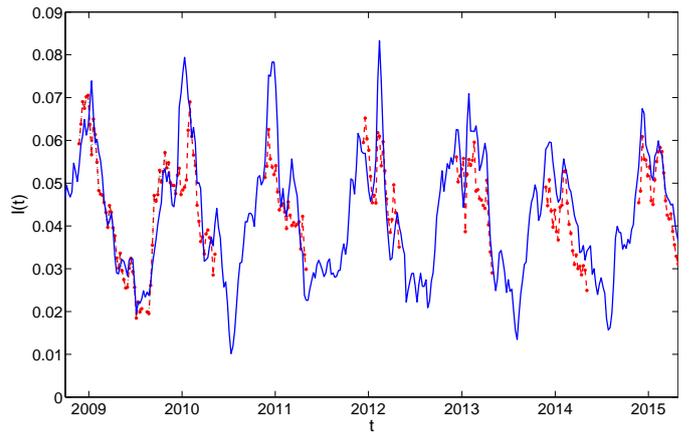}}
\caption{(Colour online) The comparison of {\tt Influenzanet}'s survey data on the common cold for the Netherlands (red curves) with the prediction obtained via the solution of linearised SIRS equations, which map the ambient air temperature over the same territory into the number of infected (blue curve, scaled to fit observations).}
\label{ccoldfig}
\end{figure}

At the same time, the applicability of the convolution (\ref{convI}) induces the inverse problem of its kinetic mechanism, which could differ from the standard compartmental approach of mathematical epidemiology applied to (\ref{eqS})--(\ref{eqR}).

The differential equation, whose solution is Eq.~(\ref{convI}) is principally based on the linearisation of Eq.~(\ref{eqI}) around the steady state $I_s$. Introducing small quantities $i$ and $s$ as $I=I_s+i$ and $S=S_s+s$, it is easy to show that this linearisation (neglecting small terms $\kappa i$, $\kappa s$ as well), leads to the equation:
\begin{equation}
\frac{di}{dt}=k_0I_s\left(s+\kappa(T(t))S_s\right).
\label{linI}
\end{equation}

Note that the steady state values (\ref{sstates}) eliminate not only constant terms. but the linear term with respect to $i$ as well. As a result, the interpretation of two terms which remain in Eq.~(\ref{linI}) can be given as follows.
The steady state value $I_s$ is a mean normal level of the infection present in a population, $k_0$ is a standard mean classic contact rate, and the term $k_0I_ss$ corresponds to the standard contact infecting  processes for a small number of people, which are instantly susceptible to one of a large variety of possible rhinoviruses. This term does not depend on the history, which is averaged due to the usage of $I_s$. On the other hand, $S_s$ is the larger part of a principally susceptible population, which has a negligible probability to catch flu (common cold) under normal everyday conditions ($\kappa=0$), without any additional stress.

At the same time, ambient temperature changes represent one such physiological stress, e.g. overcooling due to inappropriate clothing for bad weather. It has been found \cite{Makinen2009} that a $1^o$~C decrease in the ambient temperature increases the estimated statistically significant risk for common cold by $2.1\%$ in the nearest days. Thus, $\kappa(T(t))$ is not a variation of the contact rate, but the variation of probability of a depressed resistance,  and the product $\kappa(T(t))S_s$ calculates the part of the whole stable non-infected and non-recovered population, which is in danger of catching flu. At the same time, the mathematical structure of Eqs.~(\ref{eqS})-(\ref{eqR}), see Eqs.~(\ref{eqr})-(\ref{eqn}) and their consequences (\ref{eqrlin}), (\ref{linI}), have shown that the $T$-dependent term is {\it an outer excitation}. This means that a time delay is not related to the transmission. The illness emerges some days after overcooling, but one does not need to be included into some contact process during this time. 

Therefore, from the kinetic point of view, the second term in Eq.~(\ref{linI}) does not satisfy the conventional interpretation of an autocatalytic compartmental interaction. It corresponds to ``an influx into the sphere of reaction'' of ``the reagent'' with different properties.

\section{Conclusion}

The results presented in this paper demonstrate that the approach based on the SIRS compartmental model with the variable coefficient driven by the ambient temperature can adequately reproduce a certain kind of respiratory diseases, namely, common cold.

At the same time, the simplified linear equation with external forcing  that we have used, is based on a kinetic picture, which differs from the conventional interpretation of contact processes. Variation of the control parameter primarily corresponds to the variable resistance of the upper respiratory tract to rhinoviruses, which effectively depends on the air temperature but not on the changing contact rate. This  interpretation is supported by medical studies \cite{Eccles2002,Mourtzoukou2007,Makinen2009}. In addition, they show that one can neglect variations in humidity \cite{Makinen2009} for modelling of the common cold. 

The kinetic interpretation corresponding to the variable ``influx'' of overcooled persons requires a time delay between the cold exposure  and the active phase of common cold. It has been shown that  $\Delta=3$~days allows quite reasonable correspondence between the observational and the simulated data. This value agrees with the clinical observations (from 1-2 days 
\cite{Eccles2002,Makinen2009} to 3-5 days \cite{Mourtzoukou2007}) as well as other values of the used constants.

In particular, the value of the recovery time $\theta$ is chosen since it allows for reproducing both the period and the shape of the observed oscillations. At the same time, it corresponds to the duration of common cold consequences  \cite{Mourtzoukou2007} but is sufficiently smaller than the characteristic value for mutations of influenza viruses \cite{Axelsen2014}. 
The last fact is in line with the result, which indicates that the considered temperature-driven model does not properly reproduce  the curve of influenza. Moreover, the lower part of the model curve corresponds to negative values of epidemic activity, which are completely nonsensical.

Evidence of a correspondence between  seasonal variations of the ambient temperature and influenza-like illnesses suggests that the influence of temperature  should be taken into account in future, more specific, models. However, the more rigorous modelling may require more detailed, e.g. multilevel approaches, see for example \cite{Sanz2014}, considering various seasonal influenza-like diseases in their interactions.

Finally, it should be concluded that the present model is restricted to the case of the common cold.
But it provides estimations with reasonable accuracy and its basic constant parameters  are constant for a long time period. Therefore, their knowledge is adjusted to historical data and meteorological forecasts of the ambient temperature allows for forward prediction of this kind of diseases.


\end{document}